\documentstyle [12pt]{article}
\title{
\hspace{3.8truein}{\small UU-HEP/91/12}\\
{\bf Two-Loop Analysis of Non-abelian Chern-Simons Theory}}
\author{Wei Chen$^{1,2,}$\footnotemark[1]~, Gordon W. Semenoff$^{2,}$
\footnotemark[2]~ and~ Yong-Shi Wu$^{1,}$\footnotemark[1]\\ \\
       $^1$~~Department of Physics, University of Utah\\
            Salt Lake City, Utah 84112, USA\\ \\
        $^2$~~Department of Physics, University of British Columbia\\
          Vancouver, British Columbia, Canada V6T 1Z1}

 \footnotetext{\dag Work supported in part by the U.S. National
               Science Foundation under grant No. PHY-9008482.}
 \footnotetext{\ddag  Work supported in part by the Natural Sciences
               and Engineering Research Council of Canada.}
\date{}
\begin{document}
\maketitle

\vspace{0.2truein}
\begin{abstract}
Perturbative renormalization of a non-Abelian Chern-Simons gauge theory
is examined. It is demonstrated by explicit calculation
that, in the pure Chern-Simons theory, the beta-function for the
coefficient of the Chern-Simons term vanishes to three loop order.
Both dimensional regularization and regularization by introducing a
conventional Yang-Mills component in the action are used. It is shown that
dimensional regularization is not gauge invariant at two loops. A variant
of this procedure, similar to regularization by dimensional reduction
used in supersymmetric field theories is shown to obey the Slavnov-Taylor
identity to two loops and gives no renormalization of the Chern-Simons term.
Regularization with Yang-Mills term yields a finite integer-valued
renormalization of the coefficient of the Chern-Simons term at one loop,
 and we conjecture no renormalization at higher order.
We also examine the renormalization of
Chern-Simons theory coupled to matter.  We show that in the non-abelian case
the Chern-Simons gauge field as well as the matter fields require infinite
renormalization at two loops and therefore obtain nontrivial anomalous
dimensions.  We show that the beta function for the gauge coupling constant
is zero to two-loop order, consistent with the topological quantization
condition for this constant.
\end{abstract}

\renewcommand{\theequation}{\thesection.\arabic{equation}}
\baselineskip=24.0truept

\newpage
\begin{center}
\section{Introduction}
\end{center}
 \setcounter{equation}{0}
\vspace{3 pt}

There has recently been much interest in topological quantum field theories.
They are conjectured to have something to do with a high temperature
phase of quantum gravity (or superstring theory). Also of particular
 mathematical interest are certain models defined in three spacetime
dimensions,  whose actions consist purely of either (non-abelian) gauge
theory or gravitational Chern-Simons terms. These have been shown to
have an intimate connection with the classification theory of knots on three
dimensional spaces \cite{witt} and with integrable statistical
mechanics models and rational
conformal field theories in two dimensions\cite{mose}.

Formally, Chern-Simons theory is a strictly renormalizable quantum field theory
and its perturbation expansion contains logarithmic divergences.  However, the
topological nature of the theory allows only trivial, finite renormalization of
the Chern-Simons term itself.  It is interesting to verify whether this indeed
happens in the context of renormalized perturbation theory.

 There is a formal proof that the correlation functions of Wilson
 loop operators
in Chern-Simons gauge theory are topological invariants \cite{witt}.
Of particular
interest is the possibility that the perturbative expansion of these
 correlation
functions defines, order by order, new topological invariants of
 3-manifolds and
also knot and link invariants\cite{gmm}.  A necessary condition for this is
that the theory is at least perturbatively finite.  It is a further requirement
that it is invariant under gauge transformations and diffeomorphisms.

In this paper we shall examine the perturbative structure of
 Chern-Simons theory.  We shall show that a pure Chern-Simons
theory is finite to two loops and that
the beta function for the gauge coupling constant vanishes to at least three
loops.  A novel feature of our approach is the use of regularization by
dimensional reduction.   This regularization scheme renders the perturbation
theory particularly simple and gauge invariant to at least
 three loop order.  A
brief version of our  result appeared
in ref.\cite{pert}\cite{lett}.

We shall also consider the renormalization of a
theory where fermions and scalar
matter fields couple to a Chern-Simons gauge theory.  Since the matter field
actions necessarily depend on the spacetime metric, this is no longer a
topological field theory.  Infinite renormalization is necessary
 and the field operators acquire nontrivial anomalous dimensions.
However, we shall show that the beta function for the gauge coupling constant
vanishes to two loop order \cite{rcom}. This result is
consistent with the topological quantization
 condition for the gauge coupling constant.  When the gauge
coupling is the only interaction and the matter is massless the
matter-coupled Chern-Simons theory can be regarded
as a 3 dimensional conformal
field theory.

We shall devote the first part of this paper to the renormalization of
pure Chern-Simons theory where the Euclidean
action consists of the three-form
\begin{equation}
             I_{C.S.}=-\frac{i\kappa}{4\pi} \int
                 tr(A\wedge dA+\frac{2}{3} A\wedge A\wedge A)
\label{action0}
\end{equation}

Here the antihermitean generators of the Lie algebra are normalized so that
\begin{equation}
               tr(T^a T^b)=-\frac{1}{2}\delta^{ab}
\end{equation}
and the gauge field is a Lie algebra-valued one-form
\begin{equation}
               A=A^a T^a
\end{equation}

Under the gauge transformation
\begin{equation}
               A \rightarrow g^{-1}(A+d)g
\end{equation}
the integrand in (\ref{action0}) transforms like a closed form
\begin{equation}
      \delta I_{C.S.}=-\frac{i\kappa}{4\pi}
\int\left( {d(g^{-1}dg\wedge A)
                      +\frac{1}{3}
 g^{-1}dg\wedge g^{-1}dg\wedge g^{-1}dg}\right)
\label{tran1}
\end{equation}
The first term in the transformation is globally exact if $A$ obeys suitable
boundary conditions and the last term is an integer representing a winding
number of the mapping, $g(x)$, of the spacetime manifold onto the gauge
group \cite{djt}. For
any semisimple Lie group
\begin{equation}
                   \Pi_0 ({\cal G})=\Pi_3 (G)=Z
\end{equation}
where $\cal G$ denotes the local gauge group based on the Lie group G.
The gauge
transform of the action is therefore a
 constant proportional to this integer
\begin{equation}
                \delta I_{C.S.}=2\pi i\kappa \times integer
\label{tran2}
\end{equation}

To quantize this theory there are several issues which must be addressed.
The first is gauge invariance. We consider the path integral representation
of the Euclidean partition function
\begin{equation}
               Z=\int [dA] exp(-I_{C.S.}[A])
\label{z}
\end{equation}
The integrand of this formal expression is only gauge invariant under large
gauge transformations (\ref{tran2}) if the constant $\kappa$
is an integer. This is the
well-known quantization condition of the topological
 mass term \cite{shon}\cite{djt} of a
topologically massive nonabelian gauge theory in three dimensions\cite{djt}.
In order to do perturbative computations, it
is also necessary to fix a gauge.

A second issue is ultraviolet regularization
\cite{djt}\cite{pisr}\cite{pert}.  First of
all, unlike a conventional field theory, the integrand in (\ref{z}) is an
oscillating function of $A$ even in Euclidean space.
Thus, even there we do not
get a well-defined path integral with a convex Gaussian measure.  Also,
(\ref{action0}) describes a nonlinear field theory but contains no dimensional
parameters. Its perturbative expansion is therefore strictly renormalizable and
it is necessary to define perturbative calculations with a cutoff. It is also
necessary to examine the question of renormalization of the parameters, terms
in the effective action, etc.

Perturbatively (\ref{action0}) is the simplest gauge theory we know of.
In the Landau gauge (see below),
aside from the usual Faddeev-Popov ghosts it has the
antisymmetric gloun propagator
\begin{equation}
       \frac{4\pi}{\kappa} \delta^{ab}
        \frac{\epsilon_{\mu \nu \lambda}p^\lambda}{p^2}
\label{ru1}
\end{equation}
and an antisymmetric 3-gloun vertex
\begin{equation}
             i\frac{\kappa}{4\pi}f^{abc}\epsilon_{\mu \nu \lambda}
\label{ru2}
\end{equation}
These are much more elementary in form than their counterparts in QCD for
example.

In fact in an axial gauge, $n_\mu  A_\mu=0$,
 the theory is trivial -- the ghosts
decouple and the interaction term is zero. The gauge constraint is nonlinear
but this is not seen perturbatively. The action is (if $n_\mu =\delta_{\mu 0}$)
\begin{equation}
     I_{C.S.}=\frac{i\kappa}{4\pi} \int \epsilon^{0 i j} A^a_i\dot{A}^a_j
\label{action1}
\end{equation}
Perturbatively this gives the naive expectation that the theory is trivial
-- there are no interactions and the effective action is identical to the
bare action. However, the apparent triviality of the theory is naive in that
the transformation to an axial gauge is not a gauge transformation under
which the Chern-Simons term is
invariant. It cannot been done on a compact space without introducing
singularities in the gauge connection $A$.
On an open space it has nontrivial behavior at infinity
and the change in the Chern-Simons term in (\ref{tran1}) is not
proportional to a correctly quantized integer.
Therefore the theory described
by (\ref{action0}) and (\ref{action1}) are not identical.
However, their perturbative structure must be very similar.

Although the above statements about triviality of the Chern-Simons theory are
naive, in a slightly modified form they are likely true. This is a result of
a large symmetry of (\ref{action0}) --
diffeomorphism invariance. The integrand in the
action is a three-form. A three-form can be written down without reference to
any metric -- so (\ref{action0})
is invariant under any local deformation of the metric
of the spacetime manifold. If this symmetry survives at the quantum level
the only terms which could possibly appear in the effective action are those
which are covariant under general coordinate transformations and which also
do not depend on the metric. This means that the effective action is
necessarily also a local three-form containing the operators $A\wedge dA$ and
$A\wedge A\wedge A$. Furthermore the relative coefficient of these two terms
is fixed by the requirement of local gauge invariance -- the Chern-Simons
term is gauge invariant only if the two terms occur with the same relative
coefficient as in (\ref{action0}).

This means that if the symmetries of the theory under both local gauge
transformations and arbitrary variations of the
spacetime metric survive after quantization the only quantity
in (\ref{action0}) which can change under renormalization
is the overall coefficient. Furthermore, if invariance under large gauge
transformations (\ref{tran2}) remains a symmetry
it must change in such a way that $\kappa$ remains an integer.

It is impossible to fix a covariant gauge in this model without introducing a
metric. It is furthermore impossible to regulate ultraviolet divergences
without a metric -- $i.e.$ using any cutoff implies use of a distance scale
which only makes sense when there is a metric. Thus the precise definition of
the field theory in (\ref{z}) requires a metric.
The diffeomorphism invariance of
the renormalized theory would imply that the dynamics at momentum scales much
smaller than the ultraviolet cutoff are independent of the metric.

Of course, it is possible that radiative corrections
generate gauge field independent but metric dependent
terms in the effective action, an example being the
 gravitational Chern-Simons term itself \cite{witt}.  If
these terms are local they can be cancelled by adding
 local counterterms which depend only on the metric.

If there were a diffeomorphism anomaly, we also could not
exclude the possibility of
generating finite nonlocal terms in the
effective action which would depend on the metric. An example is the term
\begin{equation}
   \int F_{\mu \nu} \frac{1}{\sqrt{D_{\lambda}D^{\lambda}}} F^{\mu \nu}
\end{equation}
If there were also a gauge anomaly, more possibilities would appear.

In this paper we shall use renormalized perturbation theory to examine
whether it is in fact true that the gauge and diffeomorphism symmetries
survive quantization. We shall find that, as is usual in a gauge field theory,
the answer to this question depends on the regularization scheme used. We
shall discover that, even though this theory is power-counting strictly
renormalizable, the antisymmetric tensor structure of the vertices and gluon
propagator renders it super-renormalizable. That is, we shall argue that
once the tensor structure is taken into account and after suitable
regularization there are only a finite number of primitively divergent Feynman
diagrams.

We shall examine three types of regularization. The first introduces a Yang-
Mills term
\begin{equation}
                 I_{Y.M.}=\int -\frac{1}{2e^2}trF_{\mu \nu}F^{\mu \nu}
\label{ym}
\end{equation}
to the action. The coupling constant $e^2$ is dimensional and acts as
an ultraviolet cutoff. This makes the model power-counting super-
renormalizable at the expense of complicating the Feynman rules. There are only
a finite number of divergent Feynman diagrams which must be regulated by
independent means. We shall discuss several ways of doing this. We shall
find that with this regularization there is a finite, correctly quantized
one loop correction to the overall coefficient of the Chern-Simons action,
\begin{equation}
\kappa \rightarrow \kappa + \frac{C_2(G)}{2}~{\rm sign} (\kappa)
\label{ka}
\end{equation}
where $C_2(G)$ is the value of quadratic Casimir operator of G in the adjoint
representation. This correction has been computed before \cite{pisr}  and in
particular Witten has derived it in a nonperturbative
calculation of the phase of the
determinant of the covariance in the quadratic
approximation to the action \cite{witt}.
The dependence on the shift of $\kappa$ in (\ref{ka})
on the sign of $\kappa$ was not noticed in previous literature.
It is in fact necessary for covariance of
the theory under orientation reversing isometrics of
the manifold on which it is defined.
Even though the Chern-Simons term (\ref{action0})
does not depend on the spacetime metric
it changes sign under a coordinate transformation which reverses
the orientation of the spacetime manifold.
Thus, if the manifold has an orientation-reversing
isometry the partition function (\ref{z})
does not depend on the sign of $\kappa$.
The appearance of sign($\kappa$) in (\ref{ka})
guarantees that in the effective
action one can compensate the sign reversal of the Chern-Simons term by
reversing the sign of $\kappa$.  (rel{ka}) is
a result of our explicit calculation
and implies that the magnitude of $\kappa$ increases.

We then consider dimensional regularization.
It has the advantage that it does
not complicate the Feynman rules and higher orders in perturbation theory are
more accessible. We shall show that it gives a perfectly consistent gauge
invariant regularization of the theory at one-loop level. However it fails to
satisfy the Slavnov-Taylor identity at two loops. We attribute this to a
difficulty in the dimensional continuation of the antisymmetric tensor
$\epsilon_{\mu \nu \lambda}$.

Finally we consider renormalization by dimensional reduction
\cite{pert}. This is similar
to the dimensional continuation used to regulate supersymmetric field
theories \cite{dred}.
The tensor algebra is performed in 3 dimensions to obtain scalar integrands and
then the dimension of the
integrations are analytically continued. This procedure is shown to satisfy
the Slavnov-Taylor identities and therefore preserves the symmetries of
pure Chern-Simons
theory to at least three loops. We conjecture that with this regularization
scheme there is no renormalization to all orders in perturbation theory,
$i.e.$ in line with our naive axial gauge expectation the theory turns out
to be  perturbatively trivial in this regularization.

In many physical applications of Chern-Simons theory the Chern-Simons gauge
fields couple to either scalar or fermionic matter fields.  This coupling is
used as a way of implementing fractional statistics \cite{frst} of the
matter fields, both in the abelian
case \cite{aban} and more recently the
non-abelian case \cite{naban}.
The picture of the quantum Hall system as an incompressible fluid whose low
energy dynamics are described by an abelian \cite{hall}
\cite{frze} or non-abelian \cite{more} Chern-Simons
theory, the matter fields have a finite energy gap and the
low energy limit of the theory, i.e. the effective action for excitations far
below the energy gap could well be a physical realization of a topological
field theory.
In a finite-size system the surface states would be described by a
conformal field theory \cite{surf}.
Abelian Chern-Simons theory is also used to describe anyon superconductivity
\cite{anys}.

Matter-coupled Chern-Simons theories necessarily involve a space-time
metric and are therefore not diffeomorphism
invariant. In general, unlike pure Chern-Simons theory,
the matter-coupled theories are not exactly solvable
and it is necessary to resort to perturbative and
semiclassical approximations to explore their structure.
However, in certain cases, such as the models with
massless matter minimally coupling to Chern-Simons fields,
in which there are no dimensional parameters, the classical
actions have a smaller symmetry - scale and conformal
invariance. An interesting question is: whether this symmetry
survives after quantization? If the answer is yes,
these massless theories might be at least partially
solvable \cite{frpa}.

Experience with quantization of
{\it renormalizable} field theories indicates that it is likely
that both the fields and the gauge coupling constants,
obtain {\it infinite}
renormalizations.
Conjectures have been made in the literature about a possible
renormalization group flow \cite{poly}  of the statistics parameter
(which contains the gauge coupling constant) in a
matter-coupled abelian Chern-Simons theory.
However, it is by now well established that in {\it abelian} Chern-Simons
theory,
when the matter has a mass gap the statistics parameter obtains only finite
renormalization at one loop \cite{nise}\cite{redl}\cite{ssw}
and has no renormalization at all from higher
loops \cite{colh}\cite{ssw}.
When the matter is massless the Chern-Simons term has finite
renormalization from two loops and beyond \cite{ssw}.
This has been demonstrated by explicit calculations for
various kinds of matter fields\cite{ssw}\cite{chen}\cite{spir}
\cite{spi2}. Higher order corrections are expected.  Finiteness of the
renormalization of the abelian Chern-Simons term implies that
both the anomalous dimension of the abelian gauge field
and the $\beta$-function for the gauge coupling must vanish.
On the other hand, as we shall see
in this paper,  matter fields coupled to abelian Chern-Simons theory
need infinite wave function renormalizations and therefore
acquire non-trivial anomalous dimensions at two loops. For the abelian theory,
similar conclusions have recently been reached by \cite{kaza}.

In this paper we shall present details of a systematic two-loop investigation
of the massless and massive matter-coupled {\it non-abelian}
Chern-Simons theories.
We shall find that, in contrast to the abelian case,
both the matter and the {\it non-abelian} gauge fields
need infinite renormalization and obtain non-zero anomalous
dimensions at two loops.
However, the 2-loop $\beta$-functions for the gauge couplings
are shown to vanish due to a delicate cancellation
between the renormalization constants.
As a direct consequence, the coupling constants do not run
and these theories are scale independent. Since the calculations
and the results involve only the divergent renormalization
parts, they are independent of regularization schemes,
and are valid for the massive matter-coupled Chern-Simons
theories as well. Furthermore, for the massless theories,
the classical Callan-Symanzik equations are modified
merely by the anomalous dimensions, and then the conformal and
scale invariance survives quantization to two-loop order.

In Section 2 we discuss the general structure of the renormalization of
pure Chern-Simons theory, establish the notation and discuss  Slavnov-Taylor
identities.  In Section 3 we examine its one-loop structure. In Section 4 we
 give details of two-loop renormalization of the pure Chern-Simons theory.
In Section 5 we include the effects of matter fields to two-loop order. Section
6 contains a discussion of the results.

\vskip 0.5truein
\begin{center}
\section{Pure Chern-Simons Theory:
Slavnov-Taylor Identities, Power Counting, and Renormalization
           Constants}
\end{center}
\setcounter{equation}{0}
\vspace{3 pt}

In order to do perturbation theory we must fix the gauge. We shall
use a linear
covariant gauge
\begin{equation}
           \partial^\mu A_\mu = 0
\end{equation}
as this sort of gauge condition can be realized on a compact space. This
requires introducing Faddeev-Popov ghosts through the standard procedure.
The ghost action is
\begin{equation}
          I_g =\int tr {\partial^\mu \bar{c}D_\mu c - \frac{1}{\beta}
                 (\partial^\mu A_\mu)^2}
\end{equation}
with
\begin{eqnarray}
                      D_\mu   &=& \partial_\mu   + [A, ~~ ]\\
                      c &=&c^aT^a
\end{eqnarray}
It gives rise to the Feynman rules
\begin{equation}
              \delta^{ab}\frac{1}{p^2}
\label{ru3}
\end{equation}
for the ghost propagator,
\begin{equation}
                   if^{abc}p_\mu
\label{ru4}
\end{equation}
for the ghost-ghost-gluon vertex as usual. For each ghost loop, a minus
sign will be added to the Feynman integral.  In this work, the Landau
gauge $\beta=0$ is used exclusively.

Now we consider the ultraviolet behavior of the theory. Since the three-
and higher-point functions are ultraviolet convergent, we concentrate on
the self-energies.
With the propagators and vertices in (ref{ru1}), (ref{ru2}), (ref{ru3}),
and (ref{ru4}), naive
powercounting shows that both the gauge field and ghost self-energies
are linearly divergent. However, more careful study will show that
it is not the case \cite{ftn3}.

First of all, the ghost self-energy diagrams
with an odd number of loops are identically zero
and the ones with an even number of loops
are potentially logarithmically divergent.
The reasons are the following: Any ghost self-energy diagram with an odd
number of loops contains an odd number of antisymmetric $\epsilon$-tensors,
so that after index contractions, one and only one of them remains.
Further, by the tensor structure, the three Lorentz indices of the remaining
$\epsilon$-tensor must be contracted  with momenta. However,
there is only one external momentum in the self-energy diagram so that
the contraction gives zero. On the other hand, the diagrams with
even numbers of loops contain an even number of
$\epsilon$-tensors. The contractions among the Lorentz indices will exhaust
all $\epsilon$-tensors so that they have normal contributions to
the ghost self-energy, which takes the
form of $\tilde{\Pi}(p)p^2$. Consequently, $\tilde{\Pi}(p)$ and therefore the
ghost self-energy is potentially logarithmically divergent.

Secondly, the gluon self-energy has dimension one.  However,
the local Euclidean
invariant linearly divergent tensor $\Lambda\delta_{\mu\nu}\delta^{ab}$
is not allowed by gauge symmetry.  The allowed tensor structure is
\begin{equation}
 \Pi_e (p)(\delta_{\mu \nu} p^2 - p_\mu p_\nu)
+ \Pi_o (p)\epsilon_{\mu \nu \lambda}p^\lambda
\label{pie}
\end{equation}
Once the projection operators $\delta_{\mu \nu} p^2 - p_\mu p_\nu$  and
$\epsilon_{\mu \nu \lambda}p^\lambda$ are extracted, the
Feynman integrals contributing to $\Pi_e (p)$ are finite and to
$\Pi_o (p)$ are potentially logarithmically divergent
\cite{naive}.
Moreover, the diagrams with even numbers of loops contain odd numbers
of $\epsilon$-tensors and thus contribute only to $\Pi_o (p)$.  Diagrams
with odd numbers of loops contain even numbers of $\epsilon$-tensors and
contribute only to $\Pi_e (p)$. Thus we see that one loop and all odd
numbers of
loops are actually
convergent and  might  give corrections only to
the symmetric part of the gluon self-energy.
However, we should be aware that any non-zero contribution to
$\Pi_e (p)$ would reflect a diffeomorphism anomaly since it implies
the quantized theory is not
independent of spacetime metric any more.
On the other hand, the logarithmic divergences
may appear only in even numbers of loops.

As a regulator one may add a Yang-Mills term
(\ref{ym}) to the Chern-Simons action (\ref{action0}). Then
the bare gluon propagator (\ref{ru1}) is replaced by
\begin{equation}
      \Delta_{\mu \nu} (p) = \frac{4\pi}{\kappa} \frac{\mu}{ p^2
  (p^2+ \mu^2)} (
          \mu \epsilon_{\mu \nu \lambda} p^\lambda + \delta_{\mu \nu} p^2
             - p_\mu p_\nu)
\label{ru5}
\end{equation}
and the three-gluon vertex (\ref{ru2}) by
\begin{equation}
         i\frac{\kappa}{4\pi} \frac{1}{\mu} f^{abc}
       [\mu \epsilon_{\mu \nu \lambda}
     - (r -q)_\mu \delta_{\nu \lambda} - (q-p)_\lambda \delta_{\mu \nu}
              - (p-r)_\nu \delta_{\lambda \mu}]
\label{ru6}
\end{equation}
where the dimensional parameter $\mu = \frac {\kappa}{4\pi} e^2$ is the
topological mass of the gluon \cite{djt})
and here plays the role of a cutoff to be
taken to infinity at the end of the calculation. This regularization also
introduces a new interaction -- the four-gluon vertex
\begin{eqnarray}
            \frac{\kappa}{4\pi}\frac{1}{\mu}
      [f^{abe}f^{cde}(\delta_{\mu \lambda} \delta_{\nu \sigma}
                       - \delta_{\mu \sigma} \delta_{\nu \lambda})
      +f^{cbe}f^{ade}(\delta_{\mu \lambda} \delta_{\nu \sigma}
                       - \delta_{\mu \nu} \delta_{\sigma \lambda})
      +f^{dbe}f^{cae}(\delta_{\mu \nu} \delta_{\lambda \sigma}
    - \delta_{\mu \sigma} \delta_{\nu \lambda})]& &\nonumber\\
                                                & &
\label{ru7}
\end{eqnarray}
With finite $\mu$ the theory is
power counting super-renormalizable.
Only a finite number of diagrams are potentially ultraviolet divergent.
Moreover, the new Feynman rules have a more complicated tensor
structure, therefore they give rise to more diagrams.

Since the tensor structure of the propagators and vertices is
altered in this
renormalization scheme, the naive power-counting arguments
given above are not strictly correct.
However, we still believe that they should apply to the
divergent parts of the quantum corrections.  It does not necessarily apply to
finite parts and, as we shall see in the next Section,
the coefficient of the Chern-Simons term is renormalized
by a finite amount at
one loop in this regularization scheme.

In general the perturbative expansions of QED$_3$ and QCD$_3$
have infrared divergences.
However, three dimensional Chern-Simons gauge
theories are infrared finite (at least) in the Landau gauge.
The essential reason is that the gluon or photon
propagator (see (\ref{ru1})) behaves like $p^{-1}$,
instead of $p^{-2}$, as $p \rightarrow 0$. This property
makes the propagator
integrable at small momenta in three dimensions.
With the $F^2$
regularization the theory looks like a massive gauge field theory, which
has been shown to be free of infrared divergences \cite{djt}\cite{pisr}.

In addition to $F^2$ regularization one
may use regularization by dimensional
continuation. Namely one dimensionally continues
the loop-momentum integrals to
 non-integral dimensions to make them convergent.
The advantage of  working with
this regularization is that one does not need to
complicate the Feynman rules at
all and higher orders in perturbation theory are
more accessible. However, it is
difficult to dimensionally continue the three-index
$\epsilon$-symbol appearing
in the Chern-Simons action.

Technically, what is needed is a rule for contracting indices of the
$\epsilon$-tensors when extracting external tensor structures from a Feynman
integral or performing tensor contractions to obtain scalar integrands.  Our
procedure is to first perform all tensor contractions to obtain scalar
integrands.  Then we
define the singularity of ultraviolet divergent integrals using dimensional
continuation.

But when contracting the tensor indices of the  $\delta_{\mu\nu}$ obtained
from the contraction of two $\epsilon$-symbols,
logically there are still two
possible choices:  Is it in the continued dimensions $\omega=3-\epsilon$ or
in the physical dimensions $D=3$?
In the first case the contractions are done in $\omega = 3 - \epsilon$
dimensions, $e.g.$
\begin{eqnarray}
                  \epsilon_{\mu \sigma \eta}\epsilon^{\mu \lambda \tau}
                &=& (\delta_\sigma ^\lambda \delta_\eta ^\tau -
                     \delta_\sigma ^\tau \delta_\eta ^\lambda)
                    \Gamma (\omega - 1)\label{conv1}\\
               \delta_\sigma ^\sigma &=& \omega\label{conv2}
\end{eqnarray}
Another possibility is
performing the tenser algebra in the strict three dimensions, $i.e.$
\begin{eqnarray}
                  \epsilon_{\mu \sigma \eta}\epsilon^{\mu \lambda \tau}
                &=& (\delta_\sigma ^\lambda \delta_\eta ^\tau -
                     \delta_\sigma ^\tau \delta_\eta ^\lambda)
              \label{conv3}      \\
               \delta_\sigma ^\sigma &=& 3\label{conv4}
\end{eqnarray}
We shall call the first choice dimensional regularization and the second
regularization by dimensional reduction. The two choices do not make any
difference in the pole terms in $\omega-3$, but may lead to different finite
parts. A priori it is unclear which choice should be used; also it is unclear
whether either of them maintains gauge invariance. We shall show below by
explicit calculations that, dimensional reduction leads to gauge
invariant results in pure Chern-Simons theory at least
to three loops and in matter-coupled Chern-Simons theory
at least to two loops, but dimensional regularization violates
the Slavnov-Taylor identities satisfied
by renormalization constants in pure Chern-Simons theory at two loops.

Now let us define relevant renormalization constants and establish the
Slavnov-Taylor identities among them. With the $F^2$ regularization,
the inverse gluon propagator is
\begin{equation}
       \delta^{ab} \Delta^{-1}_{\mu \nu} (p)
            = \delta^{ab}[\frac{\kappa}{4\pi}
                          Z_A (p)\epsilon_{\mu \nu \lambda} p^\lambda
                         + Z'_A (p)( \delta_{\mu \nu} p^2 -p_\mu p_\nu)]
\end{equation}
where
\begin{eqnarray}
         Z_A (p) &=& 1+\Pi_o (p)\label{za0}\\
         Z'_A (p) &=& \frac{1}{\mu}Z''_A (p) = \frac{1}{\mu}(1+\Pi_e (p))
                    \label{za10}
\end{eqnarray}
(When regularization by dimensional continuation is used,
$Z'_A (p)$ will be identified to $\Pi_e (p)$, which has
the dimension of mass$^{-1}$, see (\ref{pie})).
Then the full gluon propagator is
\begin{equation}
      \Delta_{\mu \nu} (p) = \frac{4\pi}{\kappa} \frac{\mu}{Z''_A (p) p^2
  (p^2+ (\frac{Z_A (p)}{Z''_A (p)})^2 \mu^2)} (\frac{Z_A (p)}{Z''_A (p)}
          \mu \epsilon_{\mu \nu \lambda} p^\lambda + \delta_{\mu \nu} p^2
             - p_\mu p_\nu)
\end{equation}
Similarly, the full three-gluon vertex is
\begin{equation}
      i\frac{\kappa}{4\pi} \Gamma^{abc}_{\mu \nu \lambda}
 =  \frac{\kappa}{4\pi}f^{abc} (Z_g (p,q,r)\epsilon_{\mu \nu \lambda}
           - Z'_g (p,q,r) [(r-q)_\mu \delta_{\nu \lambda} + cyclic]+...)
\end{equation}
where
\begin{eqnarray}
               Z_g (p,q,r) &=& 1+T_g (p,q,r)\label{zg0}\\
               Z'_g (p,q,r)
&=& \frac{1}{\mu} (1+T'_g (p,q,r))\label{zg10}
\end{eqnarray}
(As above, if regularization by dimensional continuation is used,
$Z'_g$ is identified with $T'_g$.)
To identify the gluon wave function and the three-gluon vertex
renormalization functions from $Z_A (p)$, $Z'_A$, $Z_g (p)$, and $Z'_g (p)$,
we need to first remove the regularization cut-off $\mu \rightarrow \infty$.
Suppose $Z''_A (p)$ and $Z'_g (p)$ are
convergent in this limit. As we shall see
in the subsequent Sections, it is always the case; then
\begin{eqnarray}
          \Delta_{\mu \nu} (p)
&\rightarrow&  \frac {4\pi}{\kappa}
        \frac{1}{Z_A (p)} \frac{\epsilon_{\mu \nu \lambda} p^\lambda}{p^2}\\
                   \Gamma^{abc}_{\mu \nu \lambda}
 &\rightarrow& f^{abc} Z_g (p,q,r)\epsilon_{\mu \nu \lambda}
\end{eqnarray}
It shows that the renormalization functions are $Z_A (p)$ and
$Z_g (p,q,r)$. When regularization by dimensional continuation is used,
$\Pi_e$ and $T'_g$ are convergent. But any non-zero contribution
to them would imply the space-time-metric dependence of the quantized theory
and therefore an anomaly of the diffeomorphism invariance.
We shall find that $\Pi_e$ and $T'_g$ get null contributions in the
dimensional reduction  regularization scheme up to two loops.

The renormalized ghost propagator and the $\bar{c}Ac$ vertex are
\begin{eqnarray}
           \tilde{\Delta}^{ab} (p) &=& \frac{\delta^{ab}}{Z_{gh} (p)p^2}\\
              Z_{gh} (p) &=& 1 + \tilde{\Pi} (p)
\label{zgh0}
\end{eqnarray}
and
\begin{eqnarray}
              i\tilde{\Gamma}^{abc}_\lambda (p,q,r) &=&
        if^{abc} \tilde{Z}_g (p,q,r) p_\lambda\\
\tilde{Z}_g (p,q,r) &=& 1 + \tilde{\Gamma}(p,q,r)\label{zgh10}
\end{eqnarray}

An important fact we shall see in the next Sections
is that these renormalization functions are in fact finite constants,
$i.e.$ they are independent of
both cut-off and external momenta:
\begin{equation}
   Z_A (p) = Z_A, \hspace{.2truein} Z_g (p,q,r) = Z_g, \hspace{.2truein}
                 and \hspace{.08truein} so\hspace{.08truein} on
\end{equation}
This means the theory has at most a finite renormalization, which needs no
choice of a renormalization point.
As a result, the Slavnov-Taylor identity among these renormalization
functions is simply a relation of the constants
\begin{equation}
            \frac{Z_A}{Z_{gh}} = \frac{Z_g}{\tilde{Z}_g}
\label{ward}
\end{equation}
(\ref{ward}) reflects the invariance under "small" gauge transformations
and will provide a consistency check of the
regularization methods in the next Sections.

With the renormalization constants $Z_A$ and $Z_g$, the
Chern-Simons action has the form
\begin{eqnarray}
       S &=& -\frac{i\kappa}{4\pi}
           \int tr (Z_A A\wedge dA + \frac{2}{3} Z_g A\wedge A\wedge A)\\
             &=& -\frac{i\kappa_r}{4\pi}
           \int tr (A_r\wedge dA_r + \frac{2}{3}  A_r\wedge A_r\wedge A_r)
\end{eqnarray}
where we have defined the renormalized field $A_r$,
\begin{equation}
                       A_r = \frac{Z_g}{Z_A} A
\end{equation}
and the renormalized Chern-Simons coefficient $\kappa_r$,
\begin{equation}
               \kappa_r = \frac{Z_A^3}{Z_g^2} \kappa
\label{kar}
\end{equation}

Moreover, the invariance under large gauge
transformations requires the renormalized Chern-Simons coefficient,
like the bare one, to be quantized
\begin{equation}
                         \kappa_r = integer
\label{qc}
\end{equation}
We shall check in the next Sections
whether the three regularization
schemes we are considering preserve the quantization condition (\ref{qc}).

\vskip 0.5truein
\begin{center}
\section{One-Loop Structure of Pure Chern-Simons Theory}
\end{center}
\setcounter{equation}{0}
\vspace{3 pt}
\centerline{\bf A. $F^2$ Regularization}

An important one loop result
obtained with the $F^2$ regularization is a finite renormalization
of the coefficient of the Chern-Simons action by an integer,
$\kappa_r=\kappa+\frac{1}{2}C_2(G){\rm sign}(\kappa)$.
Thus the renormalized coefficient $\kappa_r$ satisfies
the topological quantization condition (\ref{qc}).
This shift can also be obtained using
Pauli-Villars regularization \cite{aglr}.  We shall
find in the next
subsection this shift of $\kappa$ is absent when we use
regularization by dimensional continuation.

$F^2$ regularization is more effective at making higher order
diagrams convergent, as shown by naive
power counting. However, the Yang-Mills term regulates only
a part of, but not all, the divergences in the one loop diagrams.
Therefore a complement is required at this level.
We shall use regularization by dimensional continuation when
necessary. We shall find that at one loop we need to use
dimensional regularization for integrals
which contain no $\epsilon$-tensors.

One-loop renormalization of topologically massive gauge theories has been
discussed in previous literature \cite{djt}\cite{pisr}\cite{ruiz}.
Here we shall review the
one-loop structure of these theories as a $F^2$ regularization of
Chern-Simons theory.  The latter
is recovered in the limit as the cutoff $\mu$ goes to infinity.
A new aspect of our discussion is the appearance of `sign($\kappa$)' in the
shift of $\kappa$.

Let us consider the one loop diagrams in Fig.1. Using $F^2$ regularization,
the Feynman rules are the altered ones, (\ref{ru5}), (\ref{ru6}),
and (\ref{ru7}) with (\ref{ru3}) and (\ref{ru4}).
First we calculate the ghost self-energy. The diagram (1.d) is
\begin{equation}
           \tilde{\Pi}^{(1)} (p)
         = \frac{4\pi}{\kappa} \frac{C_2 (G)}{2p^2}\mu
              \int \frac{d^3 k}{(2\pi)^3}
               \frac{(k.p)^2 - k^2 p^2}{k^2 (k+p)^2 (k^2 + \mu ^2)}
\label{pi}
\end{equation}
where we have used the convention for gauge group indices
\begin{equation}
              f^{adc}f^{dcb}=\delta^{ab}\frac{C_2 (G)}{2}
\end{equation}
The integral in (ref{pie}) is finite with the le
ading term proportional to $1/|\mu|$. Evaluating it
then letting $\mu$ go to $\infty$, we get
\begin{equation}
            \tilde{\Pi}^{(1)} (p) = \tilde{\Pi}^{(1)} (0)
                = - \frac{2}{3\kappa}\frac{C_2 (G)}{2}{\rm sign}(\kappa)
\end{equation}
Then we have the one loop ghost wave function renormalization constant
\begin{equation}
              Z_{gh}^{(1)}  = 1 - \frac{2}{3\kappa}\frac{c_2 (G)}{2}
{\rm sign}(\kappa)
\label{zgh}
\end{equation}

By the symmetries, the gluon self-energy $\Pi_{\mu \nu}^{(1)} (p)$ can be
separated as
\begin{equation}
           \Pi_{\mu \nu}^{(1)} (p) =
       \frac{1}{\mu}\Pi_e^{(1)} (p)(\delta_{\mu \nu} p^2 - p_\mu p_\nu)
      +\frac{\kappa}{4 \pi}\Pi_o^{(1)}(p)
       \epsilon_{\mu \nu \lambda} p_\lambda
\end{equation}
$\Pi_o^{(1)}(p)$
yields contribution from the part of the gluon loop (1.a) which has
odd number of $\epsilon$-tensors.
$\Pi_e^{(1)} (p)$ obtains contributions from
the ghost loop (1.b), the tadpole (1.c), and the part of (1.a) with even
number of $\epsilon$-tensors.
Contracting $\Pi_{\mu \nu}^{(1)} (p)$
with $\frac{\kappa}{4\pi} \epsilon_{\mu \nu \lambda} p^\lambda /2p^2$ and
$\mu \delta_{\mu \nu} /2p^2 $, we obtain $\Pi_o^{(1)} (p)$
and $\Pi_e^{(1)} (p)$
\begin{eqnarray}
        \Pi_o^{(1)} (p)&=&\frac{4 \pi}{\kappa} \frac{C_2 (G)}{2 p^2} \mu
          \int \frac{d^3 k}{(2\pi)^3}
                   \frac{(k^2 p^2 - (k.p)^2)(5k^2+5(k.p)+4p^2+2\mu^2)}
{k^2 (k+p)^2 (k^2 + \mu ^2)((k+p)^2+\mu^2)}\\
        \Pi_e^{(1)} (p)&=&-\frac{ C_2 (G)}{8 p^2} \mu
          \int \frac{d^3 k}{(2\pi)^3}
                   [\frac{N_e (p,k)}
{k^2 (k+p)^2 (k^2 + \mu ^2)((k+p)^2+\mu^2)}+ \frac{2\mu}{\pi}]
\label{pio}
\end{eqnarray}
where
\begin{eqnarray}
          N_e (p,k) &=& 6k^6+18k^4 (k.p)+20k^4 p^2+22k^2 (k.p)p^2-12(k.p)^3
                      +9k^2 p^4\nonumber\\
         & &  -7(k.p)^2 p^2 + \mu^2 [2k^4+4k^2 (k.p)
            +k^2 p^2 +(k.p)^2]
\label{pie1}
\end{eqnarray}
The integration of (\ref{pio}) is convergent. Doing it then taking
$\mu \rightarrow \infty$, we have
\begin{equation}
               \Pi_o^{(1)} (p)= \Pi_o^{(1)} (0)
                      = \frac{7}{3\kappa}\frac{C_2 (G)}{2}{\rm sign}(\kappa)
\end{equation}
The integration of (\ref{pie1}) is power-counting linearly divergent.
But the divergence belongs to the kind that violates
the gauge invariance. Therefore any regularizations
which preserve the gauge symmetry, such as regularization by
dimensional continuation, will remove it. The calculation
further indicates that when $\mu \rightarrow \infty$,
(\ref{pie1}) is finite.
This verifies the argument in the last Section that
the (one loop) gluon wave function renormalization constant is
$Z_A^{(1)}$. It is given by
\begin{equation}
             Z_A^{(1)} = 1 + \Pi_o^{(1)}
                 = 1 + \frac{7}{3\kappa} \frac{C_2 (G)}{2}{\rm sign}(\kappa)
\label{za}
\end{equation}

Now let us proceed to show that the net result of one-loop
corrections to the $\bar{c}Ac$ vertex
vanishes as $\mu \rightarrow \infty$: The one-$\epsilon$
term of (1.i) cancels against the three-$\epsilon$ term of (1.h);
The one-$\epsilon$ term of (1.h) goes to zero; The non-$\epsilon$ terms in
the two diagrams cancel each other. This is in agreement
with the general arguments
that \cite{tayl}
to any order the $\bar{c}Ac$ vertex renormalization constant is always
\begin{equation}
                  \tilde{Z}_g^{(1)} = 1
\label{zg1}
\end{equation}

To get the one loop three-gluon vertex renormalization constant,
an easy way is to exploit the Slavnov-Taylor identity (\ref{ward}).
Substituting (\ref{zgh}), (\ref{za}), and (\ref{zg1})
into (\ref{ward}), we obtain
\begin{equation}
        Z_g^{(1)} = \frac {Z_A^{(1)}} {Z_{gh}^{(1)}} \tilde{Z}_g^{(1)}
                  = 1+ \frac {3} {\kappa}\frac{C_2 (G)}{2}
                {\rm sign}(\kappa)
\label{zg}
\end{equation}
where we have used the assumption that $\kappa$ is large enough so that the
perturbative expansion makes sense.
The direct calculation of diagrams (1.e), (1.f), and (1.g)
gives the same result.

Substituting the renormalization constants $Z_A^{(1)}$ and $Z_g^{(1)}$
in (\ref{kar}), we have
\begin{equation}
            \kappa_r^{(1)} = \frac{(Z_A^{(1)})^3}{(Z_g^{(1)})^2} \kappa
                           = \kappa + \frac{1}{2}C_2(G){\rm sign}(\kappa)
\end{equation}
It implies that the renormalized Chern-Simons coefficient satisfies
the topological quantization condition if the bare one does.
That is, quantization preserves the
large gauge invariance at one loop.

\vskip 0.3truein
\centerline{\bf B. Regularization by Dimensional Continuation }

Using regularization by dimensional continuation, we have the
simpler Feynman rules, (\ref{ru1}), (\ref{ru2}), (\ref{ru3}),
and (\ref{ru4}). Further,
the diagrams with four-point vertex in Fig. 1 do not appear. Let us do a
tensor-structure analysis first.
It is easy to see that the ghost self-energy diagram (1.d) is zero because
of the $\epsilon$-tensor.

The gluon self-energy (1.a)
and (1.b) has zero contribution
to $\Pi_o$ but might contribute to $\Pi_e$.
For the same reason,
the three-gluon vertex (1.e) and (1.f) do not contribute to the
antisymmetric part of three-gluon vertex, $T_g$, but
could contribute to its symmetric part $T'_g$; On the other hand,
being diagrams with odd number of
$\epsilon$-tensors, the $\bar{c}Ac$ vertex diagrams (1.h) and (1.i) might
develop antisymmetric parts.  All of these parts, if nonvanishing, would
give rise to vertices in the effective action which violate the
diffeomorphism invariance of the theory.

However, we find that the contributions to the gluon self-energy,
the three-gluon vertex, and the $\bar{c}Ac$ vertex, respectively,
are all canceled pairwise:
\begin{eqnarray}
          (1.b) &=& - (1.a) = -\delta^{ab}\frac{C_2(G)}{2} \Lambda^{3-\omega}
   \int \frac{d^\omega k}{(2\pi)^\omega}\frac{(k+p)_\mu k_\nu
     + k_\mu (k+p)_\nu}{k^2 (k+p)^2}\label{ab} \\
          (1.i) &=& - (1.h) = i \frac{4\pi}{\kappa}T^{abc}
\int \frac{d^3 k}{(2\pi)^3}\frac{k^\lambda \epsilon_{\sigma \eta \tau}
            r^\sigma q^\tau k^\eta} {k^2 (k+q)^2(k-r)^2}\label{ih} \\
          (1.f) &=& - (1.e) = iT^{abc} \Lambda^{3-\omega}
   \int \frac{d^\omega k}{(2\pi)^\omega}
\frac{(k-r)_\mu (k+q)_\nu k_\lambda + (k+q)_\mu k_\nu (k-r)_\lambda}
                                         {k^2 (k+q)^2 (k-r)^2}\nonumber\\
                & & \label{fe}
\end{eqnarray}
where $T^{abc} = f^{ead}f^{dbg}f^{gce}$ and $r+p+q = 0$.

Thus, with dimensional regularization at one loop there is
no renormalization to
the Chern-Simons action, $i.e.$
\begin{equation}
          Z_A^{(1)} = Z_{gh}^{(1)} = Z_g^{(1)} = \tilde{Z}_g^{(1)} = 1
\end{equation}
and no contribution to the finite part of all other vertices.

\vskip 0.5truein
\begin{center}
\section{Two-Loop Structure of Pure Chern-Simons Theory}
\end{center}
\setcounter{equation}{0}
\vspace{3 pt}

In this Section we shall use both the dimensional regularization and
the dimensional reduction to compute the two-loop diagrams of Fig. 2.
We shall find that the dimensional regularization fails to obey
the Slavnov-Taylor identity and therefore breaks the gauge symmetry,
while the dimensional reduction preserves the gauge invariance and
gives no renormalization of the Chern-Simons action.

Contrary to the case at one loop, the two-loop gluon self-energy
and three-gluon vertex diagrams are anti-symmetric under exchange of
Lorentz indices.
Therefore it is possible that two-loop corrections renormalize
the Chern-Simons action.
Also, the two-loop ghost self-energy
and $\bar{c}Ac$ vertex diagrams contain even number of $\epsilon$-tensors
so that they might renormalize the gauge fixing action. Consequently,
at this order what we need to find is the possible leading
corrections to the renormalization constants $Z_A$, $Z_g$, $Z_{gh}$,
and $\tilde{Z}_g$.

To start, let us argue that the three-point
vertices vanish identically at this order, namely
\begin{equation}
                       Z_g^{(2)} = \tilde{Z}_g^{(2)} = 1
\end{equation}
We shall discuss separately the non-planar and the planar
three-legged diagrams.

First of all, the non-planar diagrams
in Fig.2 vanish individually because of
the symmetry of the group indices. To explain this point, choose one
such diagram and cut any one of its two crossing internal propagators. In
this way, we get a one-loop four-point vertex which connects with a bare
three-point vertex through a propagator. The group factor of
the one-loop four-point vertex takes the form
of $T^{abcd}=f^{eag}f^{gbh}f^{hci}f^{ide}$. It is easy to check
that $T^{abcd}$ is symmetric under the exchange of two indices which are
not neighbors, $i.e.$ $ T^{abcd}=T^{adcb}=T^{cbad}$. On the other hand,
a bare three-point vertex carries a group factor, the
structure constant $f^{abc}$, which is anti-symmetric under the exchange of
any two indices. By the structure of the diagrams, the total
group factor
for a non-planar three-legged diagram is $N^{abc}=trT^{ajcd}f^{bjd}$. The
contraction in the group indices gives
zero, therefore the non-planar diagram vanishes.

Secondly, all planar (three-legged) diagrams cancel in pairs. The
cancellations occur as a result of the cancellations at one loop.
To see this clearly,
we label the planar three-legged diagrams in Fig. 2 that are canceled
in pair by same letter in lower and upper cases,
such as (l) and (L), (m) and (M), and so on. In each of such pairs,
the two diagrams differ from each other only in a one-loop sub-diagram,
and the two one-loop sub-diagrams just form a pair that cancel against
each other, i.e. being one of the three pairs in (\ref{ab}),
(\ref{ih}), and (\ref{fe}).

We should emphasize that the cancellations mentioned above hold
without regularization ambiguity, because all two-loop three-legged
diagrams are convergent.
In other words, {\it the vanishing of the two-loop three-legged
diagrams is independent of regularization}.

Now we consider the two-point functions, Fig. (2a)--(2k).
First we note that Fig. (2d) and (2k) vanish identically, because
each contains the one-loop ghost self-energy as a sub-diagram, which
is known to vanish with dimensional regularization (see Sec. 3B).
Since the rest of the two-loop self-energy diagrams, Fig. (2a-2j),
can be paired in a way similar to the three-legged diagrams,
it seems that the above pairwise cancellation would
happen too. However we must be very careful here since the two-legged
diagrams (2a) -- (2j) are logarithmically divergent.
In regularization by dimensional continuation,
the singularity is expressed by a single pole of the Gamma
function, $\Gamma (\frac{3 - \omega}{2})$ with $\omega \rightarrow 3$.
On the other hand, depending on
the rules, the contractions of Lorentz indices might generate factors like
$3 - \omega$ when two diagrams are added. In this case, the cancellation
can be incomplete: although the pole terms are summed to zero,
a finite term may remain.

With dimensional regularization, we shall use the contraction rules
(\ref{conv1}) (\ref{conv2} to write down the Feynman integrations.
Upon contracting with
$\frac{4\pi}{\kappa}\epsilon_{\mu \nu \lambda} p^\lambda$,
the diagrams (2.a) -- (2.c) are
\begin{eqnarray}
  (2.a) &=& \frac{1}{2}\frac{\omega-1}{2}[\Gamma (\omega -1)]^4
             (\frac{4\pi}{\kappa})^2 R^{ab}\frac{1}{p^2}I_1 (\omega )\\
  (2.b) &=& -\Gamma (\omega-1)
             (\frac{4\pi}{\kappa})^2 R^{ab}\frac{1}{p^2}I_1 (\omega )\\
  (2.c) &=& \frac{1}{2}
             (\frac{4\pi}{\kappa})^2 R^{ab}\frac{1}{p^2}I_1 (\omega )
\end{eqnarray}
where the factors $\omega - 2$ and $\Gamma(\omega -1)$ in (2.a) and (2.b)
are from the contractions of $\epsilon$-tensors, and
\begin{equation}
                 R^{ab} = R\delta^{ab} = f^{ade}f^{dcg}f^{ghe}f^{hcb}
\end{equation}
and
\begin{eqnarray}
      I_1 (\omega) &=& \Lambda^{3-\omega}\int \frac{d^\omega k}{(2\pi)^\omega}
                             \frac{F(k,p)}{k^2(k+p)^2}\\
         F(k,p) &=& \int \frac{d^3 q}{(2\pi)^3}
          \frac{\epsilon_{\sigma \eta \lambda}p^\sigma k^\eta q^\lambda
                \epsilon_{\tau \theta \xi} p^\tau k^\theta q^\xi}
               {q^2(q+k)^2(q-p)^2}\label{f}
\end{eqnarray}
The integration over $q$ is convergent, giving
\begin{equation}
             F(k,p)  = \frac{1}{32}
          [k^2|p| + |k|(p^2+k.p) - |k+p|(k.p) + |p|(k.p) - |p||k||k+p|]
\label{f1}
\end{equation}
Substituted F(k,p) into $I_1(\omega)$, the integration over k gives a pole term
\begin{equation}
             \Gamma(\frac{3-\omega}{2}) \frac{p^2}{16}
     \frac {1} {(4\pi)^ {\frac{\omega}{2}}
          \Gamma(\frac{1}{2})}
      \int^1 _0 dx \frac{\sqrt{x}}{(x(1-x)p^2)^{\frac{\omega -3}{2}}}
\end{equation}
On the other hand, adding together (2.a), (2.b), and (2.c), we have a factor
\begin{equation}
  \frac{1}{2}[1 - 2 \Gamma (\omega - 1) +
 \frac{\omega - 1}{2} (\Gamma (\omega-1))^4]
\end{equation}
which annihilates the contribution of any finite
parts of the integral $I_1(\omega)$. Then using
\begin{equation}
  [1-2\Gamma (\omega -1) +\frac{\omega -1}{2} (\Gamma(\omega -1))^4]
          \Gamma(\frac{3-\omega}{2}) = -4(1 - \gamma) -1
\end{equation}
as $\omega \rightarrow 3$, where $\gamma = 0.5772...$ is the Euler constant,
we obtain
\begin{equation}
      -\frac{1}{384\pi^2}(5-4\gamma) (\frac{4\pi}{\kappa})^2 R
\end{equation}
as the contribution to $\Pi_o^{(2)}$ of the diagrams (2.a), (2.b), and (2.c).

The gluon self-energy diagrams (2.e) and (2.f) give
\begin{eqnarray}
  (2.e) &=& -
              (\frac{4\pi}{\kappa})^2
              (\frac{C_2 (G)}{2})^2\delta^{ab}\frac{1}{p^2}I_2 (\omega)\\
  (2.f) &=&  \frac{\omega - 1}{2}(\Gamma (\omega-1))^2
              (\frac{4\pi}{\kappa})^2
              (\frac{C_2 (G)}{2})^2\delta^{ab}\frac{1}{p^2}I_2 (\omega)
\end{eqnarray}
where $I_2 (\omega)$ has a pole
\begin{eqnarray}
     I_2 (\omega) &=& (\omega-1)(\Gamma(\omega -1))^2
             \Lambda^{3-\omega}\int \frac{d^\omega k}{(2\pi)^\omega}
          \frac{p^\sigma \epsilon_{\sigma \tau \lambda}k^\lambda
                (-\frac{1}{32}|k|\delta^{\tau \eta})\epsilon_{\eta \xi \theta}
                 p^\xi k^\theta}{k^2k^2(k+p)^2}\nonumber\\
               &=&-\frac{1}{32}(\omega-1)(\Gamma (\omega -1))^3
                \Lambda^{3-\omega}\int \frac{d^\omega k}{(2\pi)^\omega}
                \frac{(k.p)^2-k^2 p^2}{|k|^3(k+p)^2}\\
               &=&\frac{1}{32}\frac{1}{8\pi^2}(\omega-1)^2(\Gamma (\omega
-1))^3
                    \Gamma(\frac{3-\omega}{2})p^2
            \int ^1 _0 dx \frac{\sqrt{x}}{(x(1-x)p^2)^\frac{3-\omega}{2}}
\end{eqnarray}
(2.e) plus (2.f) gives a factor
\begin{equation}
                  -\frac{1}{2}[2-(\omega -1)(\Gamma(\omega-1))^2]
\end{equation}
With
\begin{equation}
           [2-(\omega -1)(\Gamma(\omega-1))^2]\Gamma(\frac{3-\omega}{2})
         = 2(5-4\gamma)
\end{equation}
we find that (2.e) and (2.f) contribute to $\Pi_o^{(2)}$ with
\begin{equation}
        -\frac{1}{96\pi^2}(\frac{4\pi}{\kappa})^2(\frac{C_2 (G)}{2})^2
                 (5-4\gamma)
\end{equation}
The ghost self-energy corrections are
\begin{eqnarray}
  (2.g) &=& -\Gamma (\omega - 1)
                (\frac{4\pi}{\kappa})^2 R^{ab}\frac{1}{p^2}I_1 (\omega)\\
  (2.h) &=&
                (\frac{4\pi}{\kappa})^2 R^{ab}\frac{1}{p^2}I_1 (\omega)\\
  (2.i) &=& -2
       (\frac{4\pi}{\kappa})^2(\frac{C_2 (G)}{2})^2\delta^{ab}I_2 (\omega)\\
  (2.j) &=& (\omega-1)[\Gamma (\omega - 1)]^2
       (\frac{4\pi}{\kappa})^2(\frac{C_2 (G)}{2})^2\delta^{ab}I_2 (\omega)
\end{eqnarray}
where
\begin{equation}
      I_3(\omega) = \frac{1}{(\omega-2)(\Gamma(\omega-1))^2} I_2 (\omega)
\end{equation}
A similar calculation gives
\begin{equation}
           \tilde{\Pi}^{(2)} = (\frac{4\pi}{\kappa})^2[
                  \frac{1}{96\pi^2}(1-\gamma) (\frac{4\pi}{\kappa})^2 R
                - \frac{1}{96\pi^2}(\frac{C_2 (G)}{2})^2 (5 -4\gamma )]
\end{equation}
It is remarkable that without invoking any counterterms, the
singularities cancel among diagrams, leaving only finite contributions.

Unfortunately, defining the renormalization
constants with (\ref{za0}), (\ref{zgh0}), (\ref{zg0} and (\ref{zgh10}),
we find that either the
Slavnov-Taylor identity (\ref{ward})
or the quantization condition (\ref{qc}) is
{\it not} satisfied since
$Z_A^{(2)} \neq Z_{gh}^{(2)}$ but $Z_g^{(2)} = \tilde{Z}_g^{(2)} = 1$.
This means that
dimensional regularization is not gauge invariant at two loops.

Using the dimensional reduction, let us consider the diagrams (2.a) -- (2.k)
again. We  now find complete cancellation among these diagrams.
For instance, in the Feynman integral expressions (2.a),
(2.b) and (2.c), when the $\omega$'s in the coefficients of
$I_1 (\omega)$ are taken to be three, as a result of the contractions
rules (\ref{conv3}) and (\ref{conv4}), the summation of these three
diagrams gives zero.  It is easy to see that similar cancellations occur
between (2.e) and (2.f), (2.g) and (2.h), and (2.i) and (2.j).
Consequently, the dimensional reduction gives no renormalization to the
Chern-Simons action at two loops.

For the $F^2$ regularization at two loops, instead of performing a difficult
direct calculation, we shall argue that any correction to the Chern-Simons
coefficient from two (or higher) loops will break the topological
quantization condition (\ref{qc}).
As we have found with the $F^2$ regularization
that the renormalized Chern-Simons coefficient at one loop is
\begin{equation}
                   \kappa_r = \kappa + \frac{1}{2} C_2 (G){\rm sign}(\kappa)
\end{equation}
Assume that there is a correction from the next order. It will take the form
\begin{equation}
                      B \frac{1}{\kappa}
\end{equation}
where B is some constant. If the quantization condition is to be
satisfied, $\frac{B}{\kappa}$ must be an integer for an $arbitrary$
integer $\kappa$. The only possibility is $B=0$.

\vskip 0.5truein
\begin{center}
\section{Renormalization of Chern-Simons-Matter Field Theory}
\end{center}
\setcounter{equation}{0}
\vspace{3 pt}

In this Section we study perturbative renormalization
of D=3 Chern-Simons gauge theory  coupled to
scalar and fermionic matter. We shall show that
in the non-Abelian case the coefficient of the Chern-Simons term
has infinite renormalization and that
both the matter and the non-abelian gauge fields acquire non-vanishing
anomalous dimensions at the two-loop level.
However, the 2-loop $\beta$-function of the
gauge coupling always vanishes,
indicating that scale and conformal invariance
survive quantization and infinite renormalization.

The three dimensional Euclidean actions for the coupled theories are:
\begin{eqnarray}
             S&=&S_{SC} + S_{gf} +S_{b,f}\\
S_{CS}&=&-i\int d^3x~\epsilon_{\mu\nu\lambda}
\left\{ \frac{1}{2}A_\mu^a\partial_\nu A_\lambda^a +
 \frac{g_0}{6}f^{abc}A_\mu^aA_\nu^bA_\lambda^c\right\}\\
S_{gh}&=&\int d^3x\left\{{1\over2\alpha}(\partial_\mu A_\mu^a)^2 +
       \partial_\mu {\bar c}^a
(\partial_\mu c^a+g_0 f^{abc}A^b_\mu c^c)\right\}\\
S_b&=&\int d^3x \vert\partial_\mu\phi_i+g_0T^a_{ij}
A^a_\mu\phi_j\vert^2\label{actionb}\\
S_f&=&\int d^3x \psi^{\dagger}_i\gamma_\mu
             (\partial_\mu\psi_i+g_0T^a_{ij}A^a_\mu\psi_j)
\label{actionf}
\end{eqnarray}
Here, we have modified the Chern-Simons action (\ref{action0})
 by rescaling the gauge field so that the gauge coupling constant
is $g_0=\sqrt{2\pi\over\vert\kappa\vert}$.  Gauge
invariance requires that the matter fields couple with the same constant.
$T^a=-(T^a)^{\dagger}$ are generators of the
representations of the Lie algebra of $G$ carried by the
scalar $\phi$ or by the two-component spinor $\psi$
\cite{comm};
the two by two antihermitean gamma matrices are defined as
$\gamma_\mu = i\sigma_\mu$, $\mu= 1,2,3$; and
a relativistic gauge $\partial_\mu A_\mu^a=0$ is chosen
with ${\bar c}$ and $c$ the Faddeev-Popov ghosts.
In (\ref{actionb}) and (\ref{actionf}),
we have ignored the mass terms for the
fermion and boson so that this model is invariant under
scale and three dimensional conformal transformations.

We shall further choose the Landau gauge, with
$\alpha \rightarrow 0$, where the gauge interactions
are free of infrared divergence \cite{djt}\cite{pisr}.
The Feynman rules with this gauge fixing are summarized in Fig. 3.

By power counting, it is easy to see that with the loop
diagrams, the ghost and the scalar field self-energies
are quadratic divergent, the gluon and the fermion
field self-energies and the $\phi A\phi$ and $\bar {c}Ac$
vertices linearly divergent, while
the $\phi^* A \phi A$ and the $\psi^\dagger A\psi$ vertices
logarithmically divergent.

It is well known that
the finite parts of loop corrections in general may depend on
regularization schemes but the infinite parts do not. The infinite
renormalization and the consequent results that we shall consider
in this paper will be independent of the regularization used.

We introduce renormalization constants as follows
\begin{eqnarray}
S&=&\int{\bf \{ } Z_\phi\vert\partial_\mu\phi\vert^2
               +Z_g'g{\bf [}(A_\mu\phi)^*\partial_\mu\phi
                           +(\partial_\mu\phi^*)A_\mu\phi{\bf ]}
               +(Z_g'')g^2\vert A_\mu\phi\vert^2 -{i\over 2}
                    Z_A\epsilon^{\mu\nu\lambda}
                    A^a_\mu\partial_\nu A^a_\lambda\nonumber\\
& & -{ig\over 6}Z_g\epsilon^{\mu\nu\lambda}f^{abc}A_\mu^a A_\nu^b
       A_\lambda^c+Z_{gh}\partial_\mu{\bar c}^a\partial_\mu c^a-
       {\tilde Z}_ggf^{abc}\partial_\mu {\bar c}^a
                 A_\mu^bc^c-{1\over2\gamma_R}
                (\partial_\mu A_\mu^a)^2{\bf \} }
\end{eqnarray}
where all fields are now renormalized fields.
The Slavnov-Taylor identities read
\begin{equation}
Z_g''=(Z_g')^2/Z_\phi
\label{ward2}
\end{equation}
\begin{equation}
Z_A/Z_g=Z_{gh}/{\tilde Z}_g=Z_\phi/Z_g'
\label{ward1}
\end{equation}
which must be obtained in any gauge invariant regularization scheme.

In the following we shall compute the divergent parts of the
 renormalization constants arising from the one- and two-loop
diagrams involving external or internal matter lines. For the
sake of simplicity, we shall work {\it {exclusively with regularization
 by dimensional reduction}}. To confirm the gauge invariance of
 this regularization up to two-loop order, we calculate all
 relevant renormalization constants and verify that the
Taylor-Slavnov identities (\ref{ward1}) are obeyed.

For relevant group factors we shall use the following definitions:
\begin{equation}
tr(T^aT^b) = \delta^{ab} C_1,~~ f^{acd}f^{bcd} = \delta^{ab} C_2,~~
T^aT^a = I C_3.
\end{equation}
Note that the constant $C_2$ is what we called $C_2(G)$ before, and
only $C_1$ and $C_3$ depend on the representation R of the matter
 fields. All the group factors involved in two-loop calculations
 can be expressed in terms of these constants by the following formulas:
\begin{eqnarray}
tr (T^aT^cT^cT^b) &=&  C_1C_3\delta^{ab},\\
tr(T^aT^cT^bT^c) &=& (C_1C_3 +  {1\over 2} C_1C_2) \delta^{ab},\\
tr(T^aT^dT^c)f^{bcd} &=& - {1 \over 2} C_1C_2 \delta^{ab},\\
tr(T^dT^e)f^{adc}f^{bce} &=& - C_1C_2 \delta^{ab}.
\end{eqnarray}

\vskip 0.3truein
\centerline{\bf A. One-loop Diagrams}

With regularization by dimensional reduction, it is not hard to check
explicitly that at one loop there is no contribution at all to any
 renormalization constant from the matter-gauge-field couplings.

First it is obvious that there is no one-matter-loop diagram for either
the ghost self-energy or the ghost-ghost-gluon vertex.

For the one-matter-loop correction to the gluon self-energy, shown in Figs.
 (3.a) and (3.b), one has
\begin{equation}
(-ig)^2 \delta^{ab}
C_1 \Lambda^{3-\omega} \int {d^\omega k \over (2\pi)^\omega}
            {(2k+p)_\mu (2k+p)_\nu \over k^2(k+p)^2}.
\end{equation}
It is the same in both scalar and fermion cases, since the boson-loop
 diagram in Fig. (3.b) vanishes identically. The integral involved
\begin{equation}
        I_1 = \Lambda^{3-\omega} \int {d^\omega k \over (2\pi)^\omega}
            {k_\mu k_\nu \over k^2(k+p)^2}
\end{equation}
with $\omega=3$ seems to be linearly divergent by power-counting. However,
in dimensional regularization this integral is actually finite: Introducing
 Feynman parameter $x$ in $I_1$ and integrating over $k$ we have
\begin{equation}
 I_1 = {1\over2}(\delta_{\mu\nu} - {p_\mu p_\nu \over p^2})
          \Gamma({2-\omega \over2})
         /(4\pi)^{\omega /2} \Lambda^{3-\omega}
           \int ^1_0 dx [x(1-x)p^2 ]^{(\omega - 2)/2}.
\end{equation}
This expression is finite with $\omega = 3$, since $\Gamma(-1/2) =
-2\pi^{1/2}$.
So finally we have the one-matter-loop correction to the gluon self-energy
\begin{equation}
\Pi_{\mu\nu}^{(1)} (p) = {g^2 \over 16}
 \delta^{ab} C_1 |p|(\delta_{\mu\nu} - {p_\mu p_\nu \over p^2})
\label{pi1}
\end{equation}

 For the one-matter-loop corrections (see Figs. (3.c) and
(3.d)) to the three gluon
vertex, the diagram involving the $\phi\phi AA$ vertex vanishes
identically, and the other one (with either scalar or fermion loop)
is finite, but does not contain an $\epsilon_{\mu\nu\lambda}$ factor.

As for the one-loop correction to the matter self-energy, as shown in Figs.
(3.e) and (3.f), the scalar and fermion case are different.
For the scalar self-energy,
 each diagram is separately zero, while the fermionic self-energy is finite:
\begin{equation}
-i\Sigma^{(1)}(p) = -i {g^2 \over 8} p.
\label{sig1}
\end{equation}
Similarly, the one-loop corrections to the matter-gluon vertex, Fig. (3.g)
and (3.i) are finite, while in the scalar case the diagrams, Fig. (3.h)
 and (3.j),
involving the $\phi\phi AA$ vertex vanish with regularization
by dimensional reduction.

In summary, at one-loop order all renormalization constants $Z_i$
can be chosen to be unity, $Z_i=1$, at one-loop order
\cite{ftnt}
This is in agreement with ref. \cite{djt}. In particular,
with our dimensional regularization there is no shift in the Chern-Simons
 coefficient at all and the mass squared counterterm for the scalar field
 $\delta m^2=0$, while the former is finite and the latter is linearly
 divergent in, e. g., $F^2$ regularization.

\vskip 0.3truein
\centerline{\bf B. Divergent Parts at Two Loops with Scalar Matter}

In three dimensional gauge theory,  the two-loop order is the lowest order
 in perturbation theory at which logarithmic divergences may occur. At
 this order the Feynman graphs can be classified into three sets -
 those which vanish identically before performing any integrations,
those which vanish or at least give finite results after performing one
or more of the integrals (in all cases this happens solely because of
Euclidean rotation invariance of the integral) and  those which have
 divergent parts. We confine ourselves to {\it divergent}
graphs from now on. To isolate them, certain
 results at one-loop order, such as eqs. (\ref{ab}), (\ref{ih})
and (\ref{fe}), are very useful.
These results tell us that some one-loop diagrams, either someone
alone or two of them combined together, give identically vanishing
 contribution before or after performing the momentum integration.
 When these diagrams occur as subdiagram in 2-loop diagrams, we can
first perform the associated one-loop sub-integration and then show
 either the 2-loop diagram itself or some appropriate combination
 with another 2-loop diagram will give vanishing contribution.
 The latter is similar to the cancellation between certain pair
 of diagrams that we have shown in the pure Chern-Simons case
(see Sec. 4). In this way one can eliminate many diagrams. For
 the rest, we have to carefully survey all possible diagrams
 and do power-counting one by one. Finally, we have been able
 to isolate all divergent 2-loop diagrams, as those presented
in Figs. (4.a)-(4.d).

First we note that there is only one divergent 2-loop diagram for the
 ghost self-energy, Fig. (4.b). By using (5.18) for the one-loop gluon
 self-energy insertion, we can easily obtain the divergent part
 of this diagram and, with minimal subtraction, the ghost
 wave function renormalization constant
\begin{equation}
Z_{gh}=1-{g^4\over 48\pi^2}C_1C_2{1\over3-\omega}
\label{zgh2}
\end{equation}

For the gluon  wave function renormalization $Z_A$, there are four
 divergent 2-loop diagrams involving a matter loop (see Figs.
 (4.a-1)-(4.a-4)). To extract $Z_A$ one needs to first contract each
 of them with $\epsilon_{\mu\nu\lambda}p_\lambda$. According to
 our rules for regularization by dimensional reduction, the contraction
 should be done in physical dimensions $D=3$. Then we dimensionally
 continue the integrals over loop-momenta with scalar integrands.
If one of the integrations actually leads to a finite result, it can
 be done in physical dimensions. For Fig. (4.a-1), we have
\begin{equation}
4\delta^{ab} (C_1C_3+{1\over 4}C_1C_2) {g^4 \over p^2} \Lambda^{3-\omega}
\int {d^\omega k \over (2\pi)^\omega} \int {d^3 q\over (2\pi)^3}
{k\cdot p \over k^2 q^2 (k+p+q)^2} .
\end{equation}
After performing the integration over $q$, which leads to a finite
 result in $D=3$, we are left with the following integral
\begin{equation}
p^2 I(\omega) \equiv \Lambda^{3-\omega} \int {d k^\omega \over (2\pi)^\omega}
{k\cdot p \over k^2 |k+p|}
\label{i}
\end{equation}
By introducing a Feynman parameter $x$, this integral can be
calculated to give
\begin{equation}
I(\omega) = -{1 \over 64\pi^2} \int_0^1 dx~\sqrt{x} ({1 \over \omega-3} +
\ln {\Lambda^2 \over p^2} + \ln [x(1-x)]).
\label{ii}
\end{equation}
The term containing the pole $1/(\omega-3)$ is the desired
divergent part.

Similarly, Fig. (4.a-2) leads to
\begin{equation}
8\delta^{ab} (C_1C_3+{1\over 2}C_1C_2) {g^4 \over p^2} \Lambda^{3-\omega}
\int {d^\omega k \over (2\pi)^\omega} {F(k,p) \over k^2 (k+p)^2 },
\label{4a2}
\end{equation}
where the function $F(k,p)$ is defined by eq. (\ref{f}) and is evaluated
in eq. (\ref{f1}). Substituting eq. (\ref{f1}) into (\ref{4a2}),
we find that only
 the second and third terms contribute to the divergent part with
 equal contributions, resulting in the following integral
\begin{equation}
- \Lambda^{3-\omega} \int {d^\omega k \over (2\pi)^\omega}
 {2 k\cdot p \over k^2
|k+p|} = -p^2 I(\omega).
\end{equation}
with $I(\omega)$ given by eqs. (\ref{i}) and (\ref{ii}).

Furthermore, it is easy to see that Fig. (4.a-3) leads to
\begin{equation}
-4 \delta^{ab} C_1C_2 {g^4 \over p^2} \Lambda^{3-\omega}
\int {d^\omega k \over (2\pi)^\omega} {F(k,p) \over k^2 (k+p)^2 },
\end{equation}
which exactly cancels the $C_1C_2$ term in eq. (\ref{4a2}).

Finally, Fig. (4.a-4) leads to, after inserting the one-loop result
 (\ref{pi1}) for the gluon self-energy,
\begin{equation}
- \delta^{ab} C_1C_2 {1\over 16} {g^4 \over p^2} \Lambda^{3-\omega}
\int {d^\omega k \over (2\pi)^\omega} {p^2k^2-(k\cdot p)^2 \over k^3
 (k+p)^2 }.
\end{equation}
The integral can be calculated using the Feynman parameter method which
leads to a result proportional to the integral $I(\omega)$ given
 in eqs. (\ref{i}) and (\ref{ii}):
\begin{equation}
\delta^{ab} C_1C_2 {g^4 \over 8} I(\omega).
\end{equation}
Collecting all results together, we obtain the divergent part of
$Z_A$
\begin{equation}
Z_A= 1+{g^4\over 24\pi^2}C_1C_2{1\over3-\omega}.
\label{za2}
\end{equation}
Note that the $C_1C_3$ terms are cancelled. This result is
 consistent with
the vanishing 2-loop corrections to the divergent part of
$Z_A$ in the abelian theory \cite{ssw}, where one has $C_1 =0$.

Without presenting more details, we list the divergent part of 2-loop
corrections to the three-gluon vertex, to the self-energy of the scalar
 particle and to the scalar-scalar-gluon vertex diagram by diagram,
respectively, in Appendix. They yield the divergent part for relevant
renormalization constants as follows:
\begin{eqnarray}
Z_g &=& 1+{g^4\over 16\pi^2}C_1C_2{1\over3-\omega}\label{zg2}\\
Z_{\phi} &=& 1+{g^4\over 12\pi^2} C_3({5\over2}C_3 + {13\over 8}C_2 + C_1)
                  {1\over3-\omega}\label{zphi2}\\
Z_g^{\prime} &=& 1+{g^4\over 12\pi^2}[C_3({5\over2}C_3
                 + {13\over 8}C_2 + C_1)
                 +{1\over4}C_1C_2]{1\over3-\omega}\label{zg12}
\end{eqnarray}
Furthermore we have verified that ${\tilde Z}_g$ is finite, in
 agreement with a
general theorem \cite{tayl}
that the ghost-ghost-gluon vertex is always finite and one can choose
\begin{equation}
{\tilde Z}_g = 1.
\label{zgh12}
\end{equation}
We have calculated these divergent parts of the renormalization constants
independently without using any Slavnov-Taylor identity. But it is easy to
see that they do respect the Slavnov-Taylor identities (\ref{ward1}).
This is a highly
nontrivial check that verifies that our results obtained with
regularization by
dimensional reduction are compatible with gauge invariance.

Thus, the bare and renormalized fields are related
by logarithmically divergent
renormalization. (\ref{za2}) and (\ref{zphi2})
give the corresponding anomalous dimensions:
\begin{eqnarray}
\gamma_A(g) &=& -{g^4\over 12\pi^2}C_1C_2\\
 \gamma_{\phi}(g) &=&
-{g^4\over 6\pi^2}C_3({5\over2}C_3 + {13\over 8}C_2 + C_1)
\end{eqnarray}
(These quantities, though gauge-dependent,
will appear in the anomalous scale
Slavnov-Taylor identities and in the calculation of anomalous
 dimensions for gauge
invariant composite operators.) By using eqs. (\ref{za2}) and
(\ref{zg2}), we find that a
cancellation leads to a vanishing $\beta$-function for the renormalized
coupling constant
$g=g_0 Z_A^{3/2}/Z_g$:
\begin{equation}
\beta_g (g) = 0
\label{beta}
\end{equation}
We conjecture this is true to all
orders in perturbation theory. This indicates
that the physical gauge coupling constant does not run at all, though the
theory needs infinite renormalization!
An argument similar to that in $D=4$ in
ref. \cite{schr}
 shows that (\ref{beta}) also implies the conformal invariance
for the Chern-Simons
gauge theory coupled to massless charged matter.

In passing we note a simple criterion for checking whether the $\beta$-function
vanishes or not: Since $Z_i$ are finite in pure Chern-Simons theory, the
Slavnov-Taylor identities (\ref{ward2}) and (\ref{ward1}) imply
no infinite renormalization for $g_0$
if and only
if the divergent contributions from matter loops satisfy
\begin{equation}
\delta Z_A=-2\,\,\delta Z_{gh}~~~~~{\rm or}~~~~~\delta Z_g
=-3\,\,\delta Z_{gh}
\end{equation}
Since at 2 loops only one divergent diagram contributes to
 $\delta Z_{gh}$, to
verify whether the $\beta$-function vanishes one needs to calculate
only one more renormalization constant, either $Z_A$ or $Z_g$.

\vskip 0.3truein
\centerline{\bf C. Divergent Parts at Two Loops with Fermions }

Similar results are obtained for the Chern-Simons theory coupled to massless
fermion: We find exactly the same formulas as eqs. (\ref{zgh2}),
(\ref{za2}) and (\ref{zg2}) and (\ref{zg12}) for
the divergent parts of renormalization constants in the fermion case. For
$Z_{gh}$  and $\tilde{Z}_g$, this is not too surprising, since the one-loop
correction to the gluon self-energy that they involve is the same for
the scalar or fermion loop. But for $Z_A$ or $Z_g$ this needs
some miraculous cancellations.

To see this, we note that there are four 2-loop diagrams
for gluon self-energy
 which involve a fermion loop and contribute divergent
parts to $Z_A$. See Fig
(5.a-1)-(5.a-4). Comparing them with the scalar counterparts,
Fig. (4.a-1)-(4.a-4), one
can see that only the last diagram in each case gives the same contribution,
the other three separately will not.
By contracting them with $\epsilon_{\mu\nu
\lambda} p_\lambda / 2 p^2$, we extract their contributions to
the anti-symmetric
part of the gluon propagator.
Then by calculating the trace of Dirac matrices
and first performing the integration over the
loop-momentum $q$ labeled in each
 diagram, we find following divergent contributions respectively for Fig.
(5.a-1)-(5.a-3):
\begin{equation}
- \delta^{ab} C_1C_3 {g^4 \over 2 p^2} \Lambda^{3-\omega}
\int {d^\omega k \over (2\pi)^\omega} {k\cdot p + p^2 \over |k|(k+p)^2 },
\end{equation}
\begin{equation}
 \delta^{ab} (C_1C_3 + {1\over 2} C_1C_2)
{g^4 \over 8 p^2} \Lambda^{3-\omega}
\int {d^\omega k \over (2\pi)^\omega} {k\cdot p + 2p^2 \over |k|(k+p)^2 },
\end{equation}
\begin{equation}
- \delta^{ab} C_1C_2 {g^4 \over 8 p^2} \Lambda^{3-\omega}
\int {d^\omega k \over (2\pi)^\omega} {2 k\cdot p + p^2 \over |k+p| k^2 },
\end{equation}
These integrals can be easily evaluated by introducing a Feynman parameter.
 In Appendix, we list the results diagram by diagram for $Z_A$ as well
as for $Z_\psi$.

 From Appendix one can see that the divergent parts of the renormalization
constants
$Z_A$, $Z_{gh}$ and $\tilde {Z}_g$ in the case of coupling to fermions are
the same as those (see eqs. (\ref{za2}), (\ref{zgh2}) and
(\ref{zgh12})) for the coupling to
 scalars, if they belong to the same representation $R$ of the gauge group
 $G$. By using the first Slavnov-Taylor identity in eq. (\ref{ward1})
we infer
that $Z_g$ for the fermion case should be also equal to that for the
 scalar case. The fermion wave function renormalization constant $Z_\psi$ is
\begin{equation}
Z_{\psi}=1+{g^4\over 16\pi^2} C_3(C_3+{5\over 6}C_2+{1\over 3}C_1)
           {1\over3-\omega}
\label{zpsi}
\end{equation}
Therefore, the $\beta$-function for the gauge coupling constant $g$ is
 identically zero, as in the scalar case. We did not take up the job
 of computing $Z_g^\prime$, but exploiting the second Slavnov-Taylor
 identity in eq. (\ref{ward})
we can easily obtain it from eq. (\ref{zpsi}).

\vskip 0.3truein
\centerline{\bf D.  Implications for Abelian Chern-Simons Theory}

Our results are applicable to abelian Chern-Simons
theory. For the abelian theory, we take $f^{abc}=0$ and $T^a= -i$,
then by minimal subtraction
\begin{eqnarray}
Z_A &=& 1\label{za3}\\
Z_{\phi}&=&Z_g^{\prime}=1+{7g^4\over 24\pi^2}{1\over3-\omega}\label{zphi3}\\
Z_{\psi} &=& 1+{g^4\over 12\pi^2}{1\over3-\omega}\label{zpsi3}
\end{eqnarray}
While (\ref{za3}) is known before, eqs. (\ref{zphi3}) and
(\ref{zpsi3}) are new results.

Since ultraviolet divergence of a gauge theory are
independent of whether the
coupled matter is massless or massive, our results are valid also for
{\it massive} matter, assuming there is no bare $F^2$ term.
In fact, in a
massive theory our results for $Z_i$ are the same as if one uses
Weinberg's zero-mass
renormalization scheme \cite{wein}.
 However, note that while massless matter
does not induce a finite Maxwell or Yang-Mills term,
massive matter does.
If an $F^2$ term is incorporated in the action, the theory becomes
super-renormalizable and our results can be used to derive the
leading logarithmic terms in the large gauge-boson-mass limit,
since the $F^2$
term may be viewed as a regulator for the theory without it.

\vskip 0.5truein
\begin{center}
\section{Conclusions and Discussions}
\end{center}
\setcounter{equation}{0}
\vskip 3pt

Our two-loop results have interesting and significant implications. First,
eq. (\ref{za2}) show that the non-abelian gauge fields acquire an infinite
renormalization from coupling to either massless or massive matter.
This is in contrast to the abelian case, where the coefficient of
the Chern-Simons term acquires at most finite renormalization.

The bare non-abelian coupling satisfies a topological quantization
condition: $4\pi/g_0^2=$ integer, because of the invariance
of (\ref{action0}) under large gauge transformations.
It is widely believed but not {\it a priori}
clear to us \cite{ftn2} that the same
topological quantization condition should be
respected perturbatively by the renormalized (or more precisely, physical)
coupling constant $g$. Though our vanishing $\beta$-function
is consistent with a
topological quantization condition
for $g$, the latter also requires non-renormalization
of $g$ beyond 1 loop including the finite part, which is to be verified yet
in Chern-Simons theory coupled to matter \cite{ftn3}.
It is remarkable that in all cases we have examined, abelian and non-abelian,
coupled to massless or massive matter, which can be either scalar or fermion,
the $\beta$-function for the Chern-Simons gauge coupling always
vanishes, and the `statistics parameter' $g^2_0/4\pi$ is independent of the
renormalization scale.  Thus far there seems to be
no unified understanding of this remarkable
result: The no-renormalization theorems do not apply to the
non-abelian cases, while the
topological quantization does not apply to abelian cases. We speculate that
the survival of scale and conformal Slavnov-Taylor identities is somehow
 related to the
fact that the kinetic Chern-Simons action is a topological action.
It is known that classical abelian Chern-Simons
gauge fields coupled to quantum
mechanical massive point particles are non-propagating in the sense
that, at a fixed time, they can be eliminated in terms of charged currents.
While the appearance of anomalous dimensions of field operators in the
abelian theory would seem to imply that at the quantum level the Chern-Simons
fields cease to be entirely non-dynamical when coupled to matter
fields, the fact that the gauge coupling does not run indicates that the
essential nondynamical role of transmuting the statistics
of particles survives
and is scale independent. This latter property may render these theories at
least partially solvable \cite{frpa}.

\vskip 0.5truein
\begin{center}
\section{Appendix: Divergent Two-Loop Contributions}
\end{center}
 \setcounter{equation}{0}
\vspace{3 pt}

In this Appendix we list our results, diagram by diagram, for the
divergent part of 2-loop corrections, arising from the matter coupling,
to the gluon self-energy, the three-gluon vertex,
the matter self-energy and the matter-gluon vertex.
First consider the bosonic case. The 2-loop divergent parts for $Z_A$
from the four diagrams in Fig. (4.a) are respectively
\begin{eqnarray}
 (4.a{\rm -}1) &=& {g^4\over 12\pi^2}(C_2C_3+{1\over 4}C_1C_2){1\over
3-\omega}\\
 (4.a{\rm -}2) &=& {g^4\over 12\pi^2}(C_2C_3+{1\over 2}C_1C_2){1\over
3-\omega}\\
 (4.a{\rm -}3) &=& {g^4\over 24\pi^2}C_1C_2{1\over 3-\omega}\\
 (4.a{\rm -}4) &=& -{g^4\over 48\pi^2}C_1C_2{1\over 3-\omega}
\end{eqnarray}
Summing these contributions one obtains eq. (\ref{za2}).
The four diagrams in Fig. (4.c) give rise to the following divergent
parts to $Z_g$:
\begin{eqnarray}
 (4.c{\rm -}1) &=& -{g^4\over 4\pi^2}(C_2C_3+{1\over 2}C_1C_2){1\over
3-\omega}\\
 (4.c{\rm -}2) &=& {g^4\over 32\pi^2}C_1C_2{1\over 3-\omega}\\
 (4.c{\rm -}3) &=& {g^4\over 4\pi^2}(C_2C_3+{3\over 8}C_1C_2){1\over
3-\omega}\\
 (4.c{\rm -}4) &=& {g^4\over 16\pi^2}C_1C_2{1\over 3-\omega}
\end{eqnarray}
Therefore we are led to eq. (\ref{zg2}).

 From the four 2-loop scalar propagator diagrams in Fig. (4.d) we obtain the
following divergent parts for $Z_\phi$:
\begin{eqnarray}
 (4.d{\rm -}1) &=& {g^4\over 6\pi^2}(C_3^2+{1\over 2}C_2C_3){1\over 3-\omega}\\
 (4.d{\rm -}2) &=& {g^4\over 12\pi^2}C_1C_3{1\over 3-\omega}\\
 (4.d{\rm -}3) &=& {g^4\over 24\pi^2}(C_3^2+{1\over 4}C_2C_3){1\over
3-\omega}\\
 (4.d{\rm -}4) &=& {g^4\over 24\pi^2}C_2C_3{1\over 3-\omega}
\end{eqnarray}
These contributions are summed up to give eq. (\ref{zphi2}).
Finally the eight
diagrams in Fig. (4.e) give the divergent contributions to $Z'_g$ as follows:
\begin{eqnarray}
 (4.e{\rm -}1) &=& -{g^4\over 6\pi^2}(C_3^2+{3\over 4}C_2C_3+{1\over 8}C_2^2)
{1\over 3-\omega}\\
 (4.e{\rm -}2) &=& -{g^4\over 12\pi^2}(C_1C_3+{1\over 4}C_1C_2){1\over
3-\omega}\\
 (4.e{\rm -}3) &=& -{g^4\over 24\pi^2}(C_3^2+{5\over 4}C_2C_3+{3\over 8}C_2^2)
{1\over 3-\omega}\\
 (4.e{\rm -}4) &=& {g^4\over 12\pi^2}(C_3^2+{7\over 4}C_2C_3+{5\over 8}C_2^2)
{1\over 3-\omega}\\
 (4.e{\rm -}5) &=& -{g^4\over 12\pi^2}(C_3^2+C_2C_3+{1\over 4}C_2^2)
{1\over 3-\omega}\\
 (4.e{\rm -}6) &=& {g^4\over 48\pi^2}(C_2C_3+{1\over 2}C_2^2){1\over
3-\omega}\\
 (4.e{\rm -}7) &=& -{g^4\over 24\pi^2}(C_2C_3+{1\over 4}C_2^2){1\over
3-\omega}\\
 (4.e{\rm -}8) &=& {g^4\over 192\pi^2}C_2^2{1\over 3-\omega}
\end{eqnarray}
These corrections give rise to eq. (\ref{zg12}).
We notice that the two Slavnov-
Taylor identities in eq. (\ref{ward}) are respected by the above results.

For the fermionic case, the four 2-loop diagrams that give non-vanishing
divergent contributions to $Z_A$ are shown in Fig. (5.a), with
\begin{eqnarray}
 (5.a{\rm -}1) &=& -{g^4\over 12\pi^2}C_1C_3{1\over 3-\omega}\\
 (5.a{\rm -}2) &=& {g^4\over 12\pi^2}(C_1C_3+{1\over 2}C_1C_2){1\over
3-\omega}\\
 (5.a{\rm -}3) &=& -{g^4\over 48\pi^2}C_1C_2{1\over 3-\omega}\\
 (5.a{\rm -}4) &=& -{g^4\over 48\pi^2}C_1C_2{1\over 3-\omega}
\end{eqnarray}
Their sum gives the same divergent contribution (\ref{za2}) to $Z_A$ as
the bosonic case. For $Z_\psi$, the four diagrams in Fig. (5.b)
lead to the following divergent parts:
\begin{eqnarray}
 (5.b{\rm -}1) &=& {g^4\over 48\pi^2}(C_3^2+{1\over 2}C_2C_3){1\over
3-\omega}\\
 (5.b{\rm -}2) &=& {g^4\over 24\pi^2}C_3^2{1\over 3-\omega}\\
 (5.b{\rm -}3) &=& {g^4\over 48\pi^2}C_1C_3{1\over 3-\omega}\\
 (5.b{\rm -}4) &=& {g^4\over 64\pi^2}C_2C_3{1\over 3-\omega}
\end{eqnarray}
Summing up them we obtain the result (\ref{zpsi}) for $Z_\psi$.
In the fermionic
case we did not take up the job of calculating $Z_g$ and $Z'_g$; rather we
use the Slavnov-Taylor identities to determine them. It turns out that the
2-loop divergent part of $Z_g$ arising from diagrams involving fermionic
matter is the same as the scalar case (\ref{zg2}), if the fermions and
scalars belong to the same representation of the gauge group.


\newpage
\centerline{\bf Figure Captions}
\vspace{3 pt}

Fig. 1. One-loop diagrams in pure Chern-Simons theory
        (solid line -- gluon; dashed line -- ghost)
          (a)-(c)  gluon self-energy
          (d)      ghost self-energy
          (e)-(g)  three-gluon vertex
          (h)-(i)  ghost-gluon vertex

Fig. 2. Two-loop diagrams in pure Chern-Simons theory
        (solid line -- gluon; dashed line -- ghost)
          (a)-(f)  gluon self-energy
          (g)-(k)  ghost self-energy
          (i)-(N)  planar three-gluon vertex
          (o)-(T)  planar ghost-gluon vertex
          (u)-(w)  non-planar three-gluon vertex
          (x)-(z)  non-planar ghost-gluon vertex
        Note that diagrams labelled by corresponding small and capital
        letters pairwise cancel against each other.

Fig. 3. One-loop diagrams arising from coupling to matter
        (wavy line -- gluon, dashed line -- ghost, solid line --
        bosonic or fermionic matter)
          (a)-(b) gluon self-energy
          (c)-(d) three gluon vertex
          (e)-(f) matter self-energy
          (g)-(j) matter-gluon vertex.
        Note that the diagrams (b, d, f, h, j) occur only for bosonic matter.

Fig. 4. Divergent two-loop diagrams involving bosonic matter for
        (a) $Z_A$  (b) $Z_gh$  (c) $Z_g$  (d) $Z_\phi$ (e) $Z'_g$
        (wavy line -- gluon, dashed line -- ghost, solid line --
        scalar matter)

Fig. 5. Divergent two-loop diagrams involving fermionic matter for
        (a) $Z_A$ (b) $Z_\psi$
        (wavy line -- gluon, dashed line -- ghost, solid line --
        fermionic matter). The diagram for $Z_{gh}$ looks the same as Fig. 4b.




\begin{thebibliography}{199}
\bibitem{ftn3}Note that the following arguments for the
counting of $\epsilon $-tensors are based on naive power-counting which
assumes that the regularization does not alter the structure of the
propagators and vertices.
\bibitem{naive}The reduction of the
degree of divergence by projecting out the
appropriate tensor structures was called `regularization by gauge
invariant projection' in reference (8).
\bibitem{comm}Even
though we don't distinguish these representations here,
they could be different.
\bibitem{ftnt}We therefore
     expect that in any gauge invariant regularization
   and renormalization scheme the 1-loop contribution to the
   constants $Z_i$ is finite. This is known to be the case in $F^2$
regularization (see refs (8) and (25)).
\end{thebibliography}
\end{document}